\documentclass[%
 reprint,
 amsmath,amssymb,
 aps,
]{revtex4-2}

\usepackage{graphicx}% Include figure files
\usepackage{dcolumn}% Align table columns on decimal point
\usepackage{bm}% bold math
\usepackage{placeins}
\usepackage[english]{babel}
\usepackage[fixlanguage]{babelbib}
\RequirePackage{color}\definecolor{RED}{rgb}{1,0,0}\definecolor{BLUE}{rgb}{0,0,1}\definecolor{GREEN}{rgb}{0,1,0} %DIF

\newcommand{\textapprox}{\raisebox{0.5ex}{\texttildelow}}
\newcommand{\iang}{\AA$^{-1}$}

\newcommand{\phnoo}{$\text{C}_6\text{H}_5 + \text{N}\text{O}_2$}
\newcommand{\phxno}{$\text{C}_6\text{H}_5\text{O} + \text{N}\text{O}$}
\newcommand{\nsbz}{$\text{C}_6\text{H}_5\text{N}\text{O} + \text{O}$}
\newcommand{\dsms}{$\Delta$sM($s$)}
\newcommand{\dsmst}{$\Delta$sM($s,t$)}

\begin{document}

%\preprint{APS/123-QED}

\title{Investigating dissociation pathways of nitrobenzene via mega-electron-volt ultrafast electron diffraction}

\author{Kareem Hegazy$^{1,2}$}
\author{James Cryan$^{2,3,5}$}
\author{Renkai Li${^4}$}
\author{Ming-Fu Lin${^5}$}
\author{Brian Moore${^6}$}
\author{Pedro Nunes${^6}$}
\author{Xiaozhe Shen${^5}$}
\author{Stephen Weathersby${^5}$}
\author{Jie Yang${^4}$}
\author{Xijie Wang${^5}$}
\author{Thomas Wolf$^{2,3,5}$}

\affiliation{$^1$Department of Physics, Stanford University, Stanford, California 94305, USA}
\affiliation{$^2$Stanford PULSE Institute, SLAC National Accelerator Laboratory, 2575 Sand Hill Road, Menlo Park, California 94025, USA}
\affiliation{$^3$Linac Coherent Light Source, SLAC National Accelerator Laboratory, Menlo Park, California 94025, USA}
\affiliation{\mbox{$^{4}$Department of Engineering Physics, Tsinghua University, Beijing 100084, China}}
\affiliation{$^5$SLAC National Accelerator Laboratory, 2575 Sand Hill Road, Menlo Park, California 94025, USA}
\affiliation{$^6$Diamond Light Source Ltd, Didcot, UK.}

\date{\today}% It is always \today, today,

\begin{abstract}
    As the simplest nitroaromatic compound, nitrobenzene is an interesting model system to explore the rich photochemistry of nitroaromatic compounds.
Previous measurements of nitrobenzene's photochemical dynamics have probed structural and electronic properties, which, at times, paints a convoluted and sometimes contradictory description of the photochemical landscape.
A sub-picosecond structural probe can complement previous electronic measurements and aid in determining the photochemical dynamics with less ambiguity.
%Such descriptions include sub-picosecond intersystem crossings, isomerizations through internal conversions, and dissociations from either the lowest triplet or singlet state. 
We investigate the ultrafast dynamics of nitrobenzene triggered by photoexcitation at 267~nm employing megaelectronvolt ultrafast electron diffraction with femtosecond time resolution.
We measure the first 5~ps of dynamics and, by comparing our measured results to simulation, we unambiguously distinguish the lowest singlet and triplet electronic states.
We observe ground state recovery within 160 $\pm$ 60~fs through internal conversions and without signal corresponding to photofragmentation.
Our lack of dissociation signal within the first 5~ps indicates that previously observed photofragmenation reactions take place in the vibrationally ``hot"{} ground state on timescales considerably beyond 5~ps.
\end{abstract}

\maketitle

%\tableofcontents

\section{Introduction}
\begin{figure}
    \begin{center}
        \includegraphics[scale=0.5]{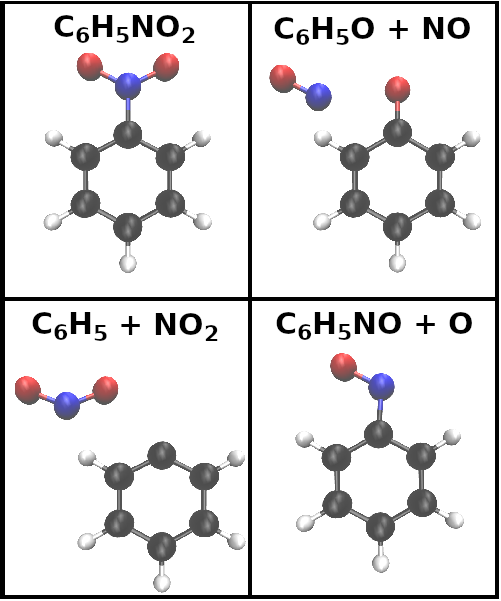}
        \caption{\label{fig:fragments}We show nitrobenzene and the primary photofragments after dissociation from a 267~nm source. Here we denote carbon, nitrogen, oxygen, and hydrogen by the colors black, blue, red, and white, respectively.}
    \end{center}
\end{figure}

Nitroaromatics exhibit an intriguing and diverse photochemistry enabled by the availability of fast intersystem crossing (ISC) channels due to the presence of the nitro functional group (NO$_2$).
Therefore, they are ideal benchmark systems for both experimental and simulation methods and have been extensively studied \cite{Lopez-Arteaga.nitro_pushpull.2013, Rodriguez-Cordoba.nitro_photochemistry.2021, Plaza-Medina.nitro_photochemistry.2011, Echeverri.nitro_polymers.2019, Zobel.nitro_ICS_picosecond.2019, He.UED_NBZ.2006, Saalbach.photoelectron.2021, schalk.NBphotoelectron.2018, Galloway.LIF.1994, Galloway.photolysisREMPI.1993, Giussani.NBZ_simulations.2017, Giussani.roaming_oxaziridineRing.2020, Hause.REMPI_roaming.2011, Quenneville.NB_simulations.2011}. 
Nitrobenzene (NB) represents the simplest nitroaromatic and is depicted in Fig.~\ref{fig:fragments}. 
Its gas-phase photochemistry following excitation in the 200-280~nm range has been investigated by various methods on timescales ranging from femtoseconds to microseconds \cite{Lin.NOrempi.2007, Galloway.photolysisREMPI.1993, Galloway.LIF.1994, Hause.REMPI_roaming.2011, Saalbach.photoelectron.2021, schalk.NBphotoelectron.2018, He.UED_NBZ.2006}. 

The three main photofragmentation pathways according to previous photolysis experiments are (1) NO$_2$, (2) NO, and (3) O abstraction (see associated fragment structures in Fig.~\ref{fig:fragments}) \cite{Galloway.photolysisREMPI.1993,Lin.NOrempi.2007}
\begin{align}
    \text{C}_6\text{H}_5\text{N}\text{O}_2 + h\nu &\rightarrow \text{C}_6\text{H}_5 + \text{N}\text{O}_2 \label{channel_NO2}\\
    \text{C}_6\text{H}_5\text{N}\text{O}_2 + h\nu &\rightarrow \text{C}_6\text{H}_5\text{O} + \text{N}\text{O} \label{channel_NO}\\
    \text{C}_6\text{H}_5\text{N}\text{O}_2 + h\nu &\rightarrow \text{C}_6\text{H}_5\text{N}\text{O} + \text{O} \label{channel_O}.
\end{align}
whereas contributions from channel \ref{channel_O} are minor. 
The branching ratio between channels \ref{channel_NO2} and \ref{channel_NO} is excitation energy dependent with NO$_2$ dominating the distribution at higher photon energies \cite{Lin.NOrempi.2007}.

Channel~\ref{channel_NO} is of particular interest, since the NO reaction product cannot result from a simple photodissociation reaction but must involve a more complicated mechanism involving both bond formation and cleavage.
Proposed mechanisms include isomerization through an oxizaridine ring \cite{He.UED_NBZ.2006, Giussani.NBZ_simulations.2017, Giussani.1-nitronaphthalene.2014, Giussani.roaming_oxaziridineRing.2020, Lin.NOrempi.2007, Xu.NBZ_simulations.2005, Suits.roaming.2008} or a roaming-type reaction \cite{Suits.roaming.2008, Hause.REMPI_roaming.2011}.

The photochemical dynamics triggered by photoexcitation at 267 nm into the lowest absorption band of NB have been investigated in a number of time-resolved studies employing different observables. 
The photolysis study by Lin \textit{et al.} finds a fragmentation timescale of 59 ns \cite{Lin.NOrempi.2007}.
Saalbach \textit{et al.} employ time-resolved photoelectron spectroscopy (TRPES) and observe dynamics in the femtosecond to picosecond range.
They assign a relaxation mechanism involving efficient ISC to the triplet manifold based on the strong spin-orbit coupling observed in NB.
However, the TRPES spectra do not provide an unambiguous signature of the ISC.
In agreement with the previous photolysis study, no evidence of photofragmentation was observed in the TRPES spectra within the first 200 ps after photoexcitation.

In contrast, a study by He \textit{et al.}~using ultrafast electron diffraction (UED) with picosecond time resolution found photofragmentation to take place within 8.8 ps \cite{He.UED_NBZ.2006}.
However, the limited temporal resolution of the study did not permit a detailed investigation of the structural dynamics leading to photofragmentation. 

Additionally, the electronic structure and photochemistry of NB was subject to several theoretical investigations \cite{Giussani.NBZ_simulations.2017, Mewes.NB_simulation.2014, Quenneville.NB_simulations.2011, Giussani.roaming_oxaziridineRing.2020, Xu.NBZ_simulations.2005, Galloway.LIF.1994, Hause.REMPI_roaming.2011}.
These studies predict reaction pathways on the potential energy surfaces of both the lowest singlet and triplet states.

With our present study, we address the disagreement in the literature with respect to the timescale of photofragmentation of NB.
We employ ultrafast electron diffraction at megaelectronvolt kinetic energies (MeV-UED), which results in femtosecond temporal resolution.
The sub-picosecond time resolution has enabled MeV-UED studies to make important contributions to the elucidation of elementary photochemical reaction mechanisms \cite{Wolf.CHD.2019, Champenois.phellandrene.2021, Liu.terpinene.2022, Yang.inelastic.2020, Yang.CF3I.2018, Wilkin.C2F4I2.2019, Liu.CH2I2_multimodal.2020, champenois_femtosecond_2023,centurion_ultrafast_2022}.
Using MeV-UED we address the question of the photofragmentation timescale of NB.
Additionally, we revisit the question about the multiplicity of the electronic state the photofragmentation takes place in.

\section{Methods}
We imaged the sub-picosecond structural dynamics of excited NB using megaelectronvolt gas-phase diffraction at the SLAC UED facility.
Reference~\cite{shen.UED_machine.2019} describes the experimental setup in detail.
Briefly, a pulsed nozzle (Parker) operating at 180~Hz delivered the gas phase NB sample to the chamber.
To deliver gaseous NB, which is a liquid at room temperature, we heated the sample to \textapprox410~K with a corresponding vapor pressure of 0.12~bar \cite{nist.NB, Brown.LiquidVapowEI.1952}.
The sample was excited at 267~nm with an intensity of $6.74\times10^{11}$~W/cm$^2$. 
The intensity dependence of the diffraction signal with respect to the pump intensity was investigated to confirm that photoexcitation was in the linear regime.
We probed the ensuing structural dynamics with a 3.7~MeV electron probe pulse.
We determined the overlap between the 267~nm pump laser pulse and the electron probe using a solid-state sample independent from the gaseous NB.
The temporal resolution of the experiment was approximately 150 fs full width at half maximum (FWHM) \cite{Wolf.CHD.2019}.
Our measurement focused on the first 1.1~ps of the pump-probe delay and a single measurement at a 5~ps delay.
Within the first 1.1~ps, we sampled pump-probe delays below 400~fs with 67~fs temporal intervals, and for delays beyond 400~fs we sampled with 100~fs intervals.
The maximum momentum-transfer of diffracted electrons was limited by signal-to-noise to 8~\iang.
Due to a hole in the detector that transmits undiffracted electrons and contamination of the scattering angles close to the edge of the hole, the minimum momentum-transfer was 2~\iang.

%NOTE: Throttle at 55, which corresponds to earlier 53 at 109 uJ at the beginning of the experiment and 44/95 uJ inside/outside the chamber measured at the end of the experiment

Diffraction patterns were simulated within the independent atom model using a publicly available code \cite{Wolf_diff} based on form factors evaluated with the ELSEPA program package \cite{Salvat2005}.
Structures of expected dissociative states were calculated by solving for the ground state structures of each fragment using the Firefly package~\cite{Granovsky.Firefly.2022}.
The ground state NB structure was taken from Ref.~\cite{Giussani.NBZ_simulations.2017}.
For simulations of the vibrationally ``hot"{} ground state and triplet state, all quantum chemical calculations were performed in the ORCA 5.0 program package \cite{Neese2012,Neese2018}.
Stationary points and vibrational frequencies were performed using B3LYP/def2-SVP \cite{Becke1993,Stephens1994,Weigend2005,Weigend2006}.
For the structural signatures, we adapted a procedure, which was previously used for spectroscopy applications \cite{Wolf2019,Wolf2021}, for electron diffraction signatures.
In short, we approximate the distribution of the photoabsorbed energy over the vibrational degrees of freedom as statistical, which can be in turn approximated by a canonical ensemble of a specific vibrational temperature using evaluated vibrational frequencies.
In the case of the electronic ground state, the vibrational energy is identical with the excitation photon energy of 4.65~eV, which corresponds to a vibrational temperature of 2940~K.
For the triplet state, we evaluate the energy difference between the ground state minimum and the triplet state minimum using CCSD/aug-cc-pVDZ, \cite{Dunning1989,Kendall1992} correct it by the zero point vibrational energies of both states, and subtract it from the absorbed photon energy.
The remaining excess vibrational energy is 2.25~eV, which corresponds to a vibrational temperature of 1335~K.
We use these temperatures to simulate molecular dynamics trajectories for both states using PBE0/def2-SVP \cite{Adamo1999}.
The starting geometries of the trajectories are the respective minimum geometries.
The trajectories are propagated for 10~ps.
To evaluate diffraction signatures, 5000 geometries are randomly sampled from the last quarter of the trajectory.

When working with the data, we use the modified molecular isotropic scattering pattern, which removes the strong $s$ dependence in the atomic scattering factors~\cite{Srinivasan.UED_overview.2003}.
\begin{equation}
    \text{sM}(s,t) = s\frac{\int I(s, \theta, t) \sin\theta d\theta }{\sum_i |f_i(s)|^2}.
\end{equation}
Here, $I(s, \theta, t)$ is the measured diffraction pattern, $\theta$ is the azimuthal angle on the detector, and $f_i(s)$ is the atomic scattering amplitude of the $i^{\text{th}}$ atom calculated with the ELSEPA program package \cite{Salvat2005}.
We highlight the molecular dynamics through the difference in modified diffraction intensity (\dsmst), by subtracting the $\text{sM}(s,t)$ before the arrival of the 267~nm pump pulse.
Taking this difference also removes the constant atomic scattering and most time-invariant detector artifacts from our measurement.

In the following analysis, we examine the rise of the observed signal and the subsequent molecular state NB occupies.
Our method and detailed results for extracting the rise time are given in Section~\ref{sc:rise_time} and briefly below.
For each time point, we sum over the absolute value of the ratio between \dsms{} and its standard deviation: $\sum_s |\Delta \text{sM}(s)/\sigma_{\Delta \text{sM}(s)}|$.
The rise time is determined by fitting these dynamics to an error function. 

\section{Results}
\begin{figure}
    \centering
    \includegraphics[scale=0.33]{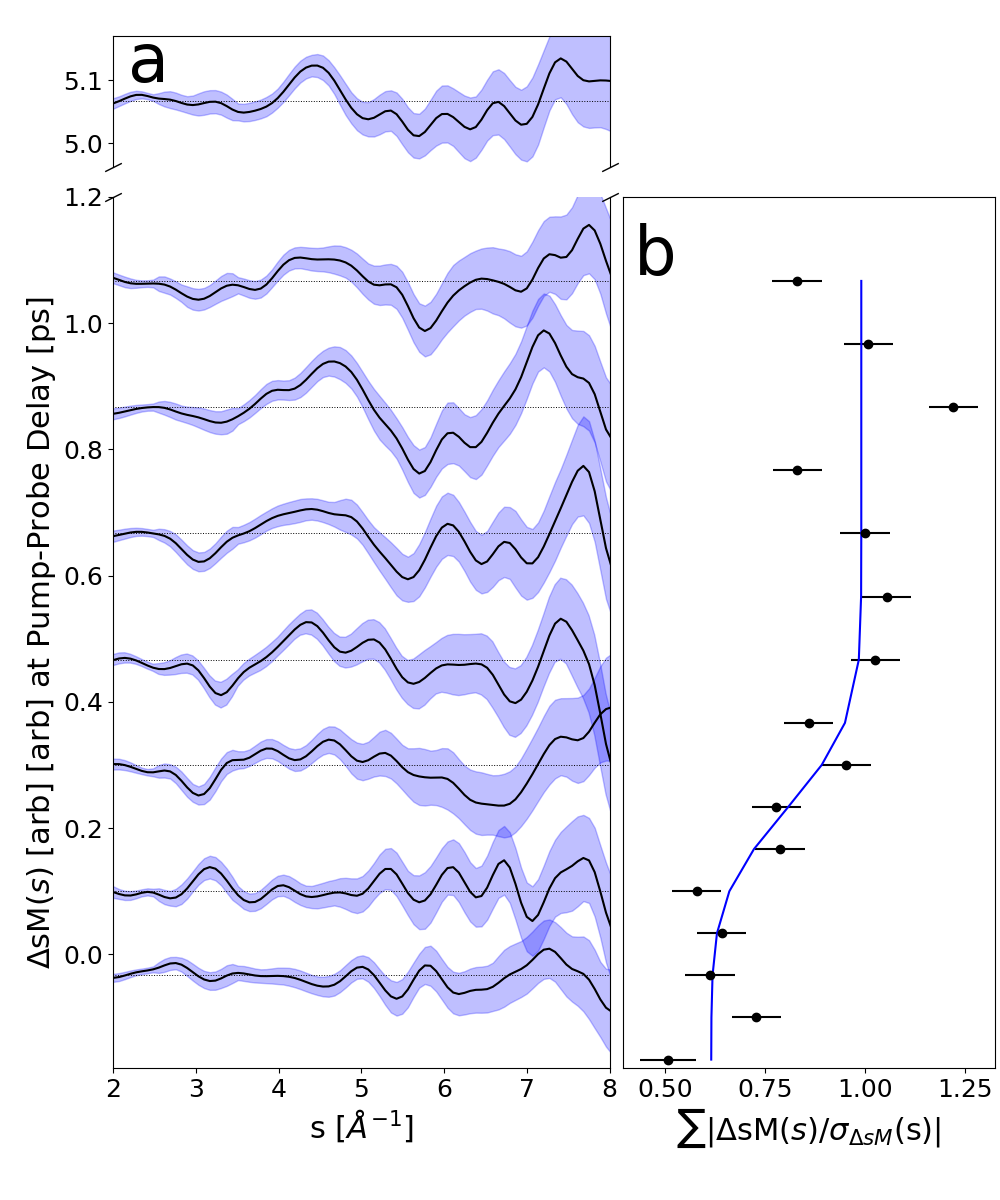}
    \caption{We show the measured time-dependent diffraction signal \dsmst{} as a function of delay in panel a. The blue bands signify one standard error of the mean. In panel b we show the rise in the signal amplitude (points) and the rise time fit (blue line). The black error bars also signify one standard error of the mean.}
    \label{fig:sms_td}
\end{figure}

Nitrobenze's absorption spectrum between 215--285~nm is dominated by a strong absorption band associated with excitation to S$_4$ and centered at 248~nm \cite{Lin.NOrempi.2007,Marshall.NB_spectra.1992,Nagakura.NB_spectra.1964}. 
It exhibits a weak shoulder associated with a transition to  S$_3$ and centered at 280~nm \cite{Lin.NOrempi.2007,Nagakura.NB_spectra.1964}.
Our 267~nm pump pulse is redshifted from the stronger S$_4$ and blueshifted from the much weaker S$_3$ band.
Thus, we primarily photoexcite NB into the S$_4$ state, but the simultaneous preparation of minor amounts of population in the S$_3$ state cannot be ignored.

Following photoexcitation to the S$_3$ and S$_4$ states, we observe the onset of a transient diffraction signal, shown in Fig.~\ref{fig:sms_td}, around time zero.
This signal is characterized by intensity increases between $4 < s < 5$~\iang{} and $7 < s < 8$~\iang, as well as a depletion between $5 < s < 6$~\iang.
We quantify the rise time of this transient signal to be $160\pm60$~fs (Fig~\ref{fig:sms_td}b) by an error function fit.
We do not believe this result is due to our instrument response as previous PES by Saalbach \textit{et al.} are sensitive to sub-100~fs lifetimes and observe a similar $195 \pm 20$~fs lifetime \cite{Saalbach.photoelectron.2021}. 

\begin{figure}
    \centering
    \includegraphics[scale=0.34]{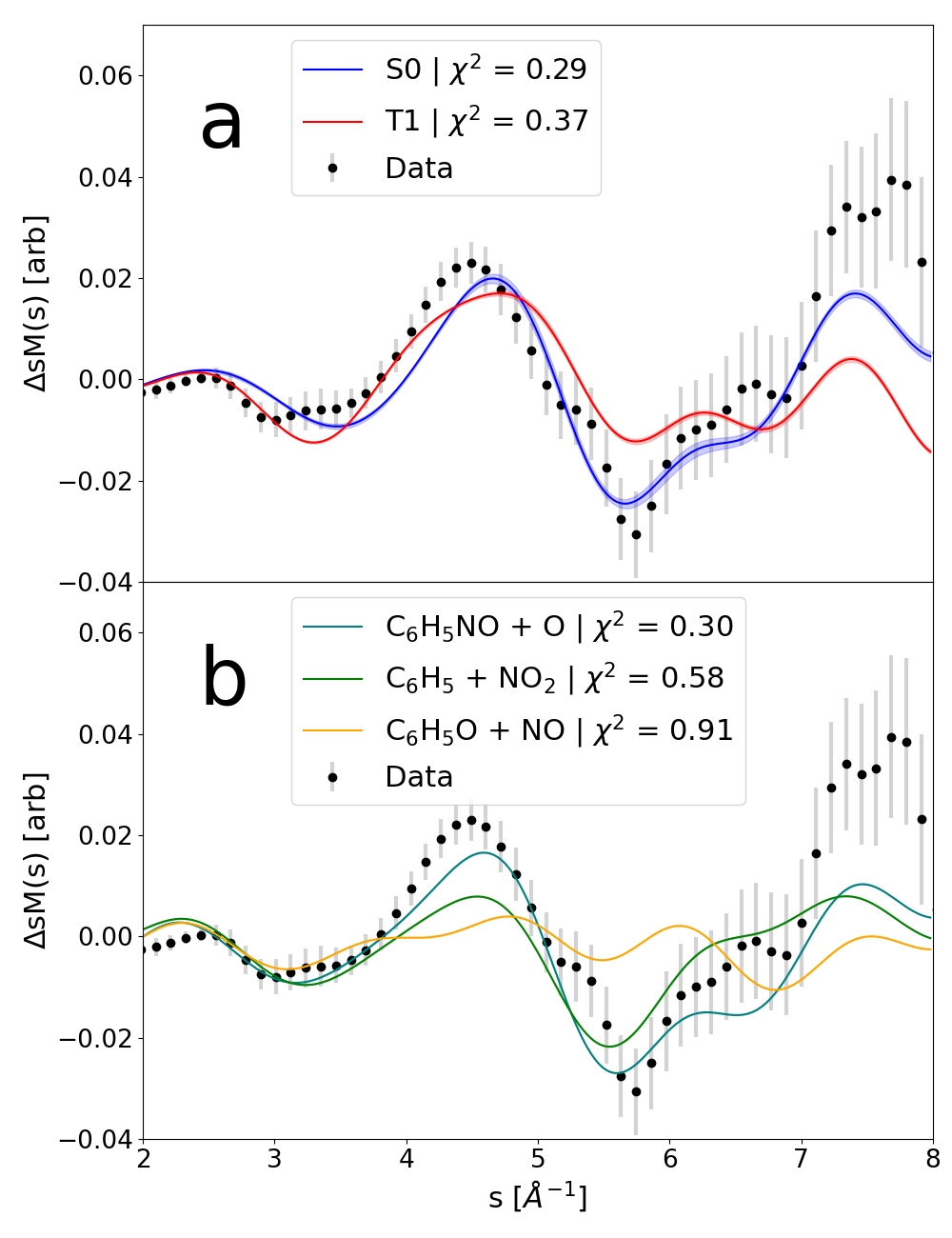}
    \caption{We show the comparison between the measured average \dsms{} signal after 700~fs when the signal has equilibrated. The error bars correspond to the standard error of the mean. Panel a shows the best fits from the vibrationally ``hot" S$_0$ and T$_1$ states. The shaded portions correspond to the standard error of the mean based on a selection of 5000 structures within these ``hot" ensembles. Panel b shows the best fits from previously observed dissociative states.}
    \label{fig:final_state}
\end{figure}

Following the onset of the transient signal at time zero we do not observe significant structural dynamics in Fig.~\ref{fig:sms_td} (panels a and b).
We consider NB has reached a steady state after undergoing either a photochemical reaction or a redistribution of energy into vibrational degrees of freedom.
To determine the structural changes associated with the onset of the  steady-state signal in Fig.~\ref{fig:sms_td}a we compare the averaged \dsmst{} between 0.75--5.25~ps, where we no longer observe changes, to simulations of different processes in Fig.~\ref{fig:final_state}.

\begin{table}[!htbp]
    \centering
    \begin{tabular}{|c|c|}
        \hline
        \textbf{Simulation} & $\boldsymbol{\chi}^2$ \\
        \hline
        S$_0$ & \hspace{0.4cm}0.29\hspace{0.4cm} \\
        \nsbz & 0.30 \\
        T$_1$ & 0.37 \\
        \phnoo & 0.58 \\
        \phxno & 0.91 \\
        \hline
    \end{tabular}
    \caption{We provide the $\chi^2$ values from fitting the simulated responses to the measured \dsms{} signal.}
    \label{tab:chisq}
\end{table}

We employ two types of simulated \dsms{} in this comparison, those based on the vibrationally ``hot"{} molecules in the S$_0$ and T$_1$ states as well as optimized product structures from the three photofragmentation channels (Fig.~\ref{fig:final_state}) \cite{Galloway.photolysisREMPI.1993,Lin.NOrempi.2007}.
In the former case (Fig.~\ref{fig:final_state}a), the \dsms{} signal originates from the substantial amount of energy, absorbed from the pump photon, being redistributed into the vibrational degrees of freedom (see the methods section).
In the latter case (Fig.~\ref{fig:final_state}b), we expect the signal to be dominated by the elimination of many pairwise distances and can therefore be approximated by a single fragment structure. 

In the comparison in Fig.~\ref{fig:final_state}a, the simulation of the vibrationally "hot"{} S$_0$ signature clearly agrees better with the experiment, specifically in the regime above $s>5.5$~\iang.
In the comparison in Fig.~\ref{fig:final_state}b, the signature of the \nsbz{} channel clearly shows the best agreement.

We quantify the agreement between experimental and simulated \dsms{} signatures by the $\chi^2$ values from linear fits of the simulations to the experimental data.
These values are listed in Table~\ref{tab:chisq}.
The vibrationally ``hot"{} S$_0$ simulation yields both the best qualitative and quantitative ($\chi^2=0.29$) agreement with the measurement.
The $\chi^2$ value of the \nsbz{} channel is slightly larger and the $\chi^2$ values of the \phnoo{} and \phxno{} channels $\chi^2$ values two and three times larger, respectively. 

\section{Discussion}
According to our fit results, the ``hot"{} ground state and the \nsbz{} dissociation channel signatures describe the experimental \dsms{} signal almost equally well.
However, the \nsbz{} dissociation was previously observed to be a minor channel \cite{Lin.NOrempi.2007}.
Thus, the corresponding experimental signature must represent a weighted superposition of \nsbz{} with different signatures of higher weight, e.g. other photofragmentation channels or vibrationally ``hot"{} signatures. 
However, fits of such superpositions to the experimental signature either yield weights incompatible with \nsbz{} being a small channel or trivial results (superposition with ``hot"{} S$_0$ signature). 
Thus, our time-resolved structural probe of NB indicates the molecule relaxes from the S$_3$ and S$_4$ states to the vibrationally ``hot"{} S$_0$ state, with a $160\pm60$~fs rise time.
This is in agreement with Saalbach \textit{et al.} who do not observe time-dependent fragmentation, from ion yields, within the first 200~ps \cite{Saalbach.photoelectron.2021}.

The observed timescale is consistent with the previous TRPES work by Saalbach \textit{et al.} which observed NB relaxation from S$_3$ and S$_4$ with a $195\pm20$~fs lifetime \cite{Saalbach.photoelectron.2021}.
Saalbach \textit{et al.} interpret the lifetime differently, as relaxation to the ``hot"{} T$_1$, but admit that their probe is not particularly sensitive to the multiplicity of the final state of the relaxation.
Previous theoretical works \cite{Giussani.NBZ_simulations.2017, Giussani.1-nitronaphthalene.2014, Mewes.NB_simulation.2014, Quenneville.NB_simulations.2011} propose relaxation channels leading to both S$_0$ and T$_1$.

Our results disagree with this scenario in two ways.
Firstly, fits of the experimental signature with a superposition of the vibrationally ``hot"{} S$_0$ and T$_1$ simulations yield either comparable or significantly larger $\chi^2$ values than with S$_0$ alone.
This can be seen in Fig.~\ref{fig:final_state}a where adding the T$_1$ signature to S$_0$ deviates the S$_0$ signature away from the data except within the small regions between 2.5-3~\iang{} and 3.5-4.25~\iang.
That is, we would expect such a superposition to improve the fit within a \textapprox1.25~\iang{} range while worsening the fit in the remaining \textapprox4.75~\iang{} fit range.
Secondly, despite a growing body of evidence for the existence of ultrafast, even sub-picosecond ISC in molecules exclusively comprised of light elements \cite{Penfold2018}, the assignment of the observed $<$200~fs timescale to ISC would require strong additional evidence.
However, our analysis of the measured time-dependent diffraction signal aided by simulated signals from thermalized molecular dynamic simulations clearly points to population of the vibrationally ``hot"{} S$_0$ via internal conversion.

The observation of the repopulation of S$_0$ and the absence of additional structural dynamics within the first 5~ps after excitation indicates, in agreement with previous studies \cite{Lin.NOrempi.2007, Saalbach.photoelectron.2021,Galloway.photolysisREMPI.1993}, that photofragmentation reactions must occur on larger timescales. 
These results are, however, in disagreement with the previous picosecond time-resolved UED study, which reported \phxno{} photofragmentation with a time constant of 8.8~ps \cite{He.UED_NBZ.2006}.
The previously observed \dsms{} signature is in reasonable agreement with our observations.
Thus, we hypothesize that the previously observed time constant represents the experimental time resolution (6~ps \cite{He.UED_NBZ.2006}) rather than the reaction dynamics of the molecule. 

The disagreement in the interpretation of the \dsms{} signals might lie in the employed models.
He \textit{et al.} simulated the signature of the vibrationally ``hot"{} S$_0$ under the assumption that it can be described by a single geometry through the incorporation of effects from the vibrational excitation \cite{He.UED_NBZ.2006}.
However, a statistical distribution of the absorbed photon energy in the vibrational modes corresponds to a vibrational temperature of 2940~K (see methods). 
It is likely that at this high vibrational temperature the experimental signature is not well described anymore by a single geometry and, therefore, the employed model is not valid anymore.
An \textit{ab initio} trajectory explicitly exploring the phase space, as performed in the present study and likely computationally infeasible at the time of the previous study, can easily overcome this limitation. 

\section{Conclusion}
We investigated the complex photochemistry of the simplest nitroaromatic, nitrobenzene, in the first sub-picosecond resolved study with specific sensitivity to the structural evolution of the compound, using MeV-UED. 
Our results reveal, aided by simulations, that after excitation to the S$_4$ and S$_3$ excited states at 267~nm, NB relaxes to the vibrationally ``hot"{} S$_0$ state within $160\pm60$~fs.
Due to the global sensitivity of MeV-UED to structural changes, we preclude previously proposed channels (e.g. relaxation via ISC to the triplet manifold and photofragmentation) from receiving a significant amount of population within the observed 5~ps after photoexcitation.
Our results point to the previously observed photochemistry happening on longer timescales through statistical mechanisms on the S$_0$ potential energy surface. 
Thus, our results present an important contribution to the elucidation of photochemical reaction mechanisms in nitrobenzene and nitroaromatic compounds in general.

\begin{acknowledgements}
We thank the 2018 SLAC MeV-UED gas-phase team and the ASTA team for their help in running our MeV-UED experiment.
We also thank our funding agencies for supporting this work.
This work is supported by the U.S. Department of Energy Office of Science, Basic Energy Sciences, Chemical Sciences, Geosciences, and Biosciences Division.
The UED work was performed at SLAC MeV-UED, which is supported in part by the Department of Energy Basic Energy Sciences' Scientific User Facility Division Accelerator and Detector research and development program, the LCLS Facility, and SLAC under contract Nos. DE-AC02-05-CH11231 and DE-AC02-76SF00515.
\end{acknowledgements}

%\appendix

%\nocite{*}
\selectbiblanguage{english}
\bibliography{nitrobenzene.bib}% Produces the bibliography via BibTeX.

%%%%%%%%%% Merge with supplemental materials %%%%%%%%%%
%TC:ignore
\pagebreak
\clearpage
\widetext
\begin{center}
\textbf{\large Supplemental Materials}
\end{center}
%TC:endignore
%%%%%%%%%% Merge with supplemental materials %%%%%%%%%%
%%%%%%%%%% Prefix a "S" to all equations, figures, tables and reset the counter %%%%%%%%%%
\setcounter{equation}{0}
\setcounter{figure}{0}
\setcounter{table}{0}
\setcounter{page}{1}
\makeatletter
\renewcommand{\theequation}{S\arabic{equation}}
\renewcommand{\thefigure}{S\arabic{figure}}
\renewcommand{\bibnumfmt}[1]{[S#1]}
%\renewcommand{\citenumfont}[1]{S#1}
%%%%%%%%%% Prefix a "S" to all equations, figures, tables and reset the counter %%%%%%%%%%

\section{Static Nitrobenzene Signal}
\begin{figure}[!h]
    \centering
    \includegraphics[scale=0.33]{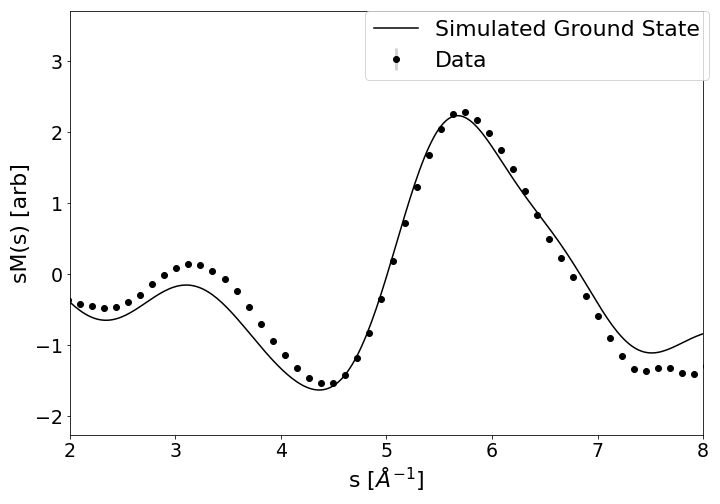}
    \caption{We show the measured static signal after subtracting the atomic scattering and measurement background. The black line corresponds to the best fit of the calculated ground state of nitrobenzene to the data.}
    \label{fig:static_SMS}
\end{figure}

\FloatBarrier
To validate the delivery of ground state nitrobenzene in the gas phase, we recorded static diffraction images taken without the 267~nm pump pulse.
To remove the atomic scattering and contributions from the experimental response function we fit the measurement using the five expected zero-crossings, from simulation.
These zero-crossings were found at 1.76, 4.99, 6.82, 9.01, and 11.74~\iang.
Each zero-crossing was fit with an exponential to guarantee that the signal at these points are equal to 0, as expected by the ground state simulation.
After subtracting these exponentials, the results were fitted with the simulated ground state molecular diffraction signal (sM($s$)).
Figure~\ref{fig:static_SMS} shows the delivered sample is in good agreement with ground state nitrobenzene and that the sample was not damaged upon delivery.

\section{\label{sc:rise_time}Rise Time Measurement}
We isolate the temporal dynamics from the two-dimensional \dsmst{} by summing over $s$. 
However, the \dsmst{} amplifies the signal at higher $s$ that suffers from poor signal-to-noise, while the raw diffraction intensity is dominated by low $s$ contributions.
To account for the strong $s^{-4}$ dependence in the raw diffraction intensity and weight data points based on their signal to noise, we normalize the \dsmst{} by its standard deviation $\sigma_{\Delta \text{sM}}(s,t)$.
\begin{equation}
    R(t) = \sum_s \left| \frac{\Delta \text{sM}(s)}{\sigma_{\Delta \text{sM}}(s)} \right|
\end{equation}
This temporal signal $R(t)$ is shown as data points in Fig.~\ref{fig:sms_td}b.

\begin{figure}
    \centering
    \includegraphics[scale=0.5]{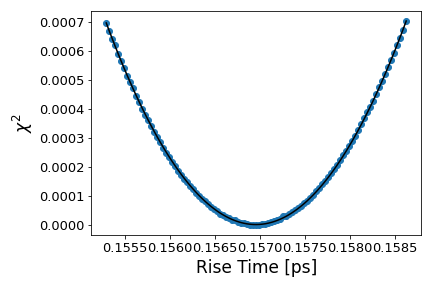}
    \caption{We show the $\chi^2(t)$ distribution and its parabolic fit that is used to retrieve the optimal rise time and its uncertainty.}
    \label{ap:fig:risetime_error}
\end{figure}

To retrieve the rise time, we fit $R(t)$ to an error function.
We simultaneously fit the mean, standard deviation, and amplitude of the Gaussian that is integrated within the error function.
The rise time corresponds to the fitted standard deviation.
To further resolve the optimal rise time and calculate its error, we fit the mean and amplitude of the error function while varying the standard deviation around the region of the first fit.
This produces a quadratic distribution of $\chi^2(t)$ values, shown in Fig~\ref{ap:fig:risetime_error}, where the minimum corresponds to the optimal rise time and the width corresponds to its error \cite{Cumpson.uncertainty.1992}.
We retrieved a rise time of $160 \pm 60$~fs.

%\section{\label{sc:diss_fit}Observing Photofragmentation within 1~ps}
%\input sections/sm_dissociation

\end{document}